\documentclass[twocolumn,nofootinbib,
showpacs,pra,aps,floatfix]{revtex4}
\usepackage{mathtext}
\usepackage{amsfonts}

\usepackage{mathrsfs}
\usepackage{bm}
\usepackage{graphicx}
\begin{document}
\title{Decay by tunneling of Bosonic and Fermionic Tonks-Girardeau Gases}

\author{A. del Campo}
\email{qfbdeeca@ehu.es} \affiliation{Departamento de Qu\'\i mica-F\'\i sica, Universidad del Pa\'\i s Vasco, Apdo. 644, 48080 Bilbao, Spain}
\author{F. Delgado}
\email{qfbdeacf@ehu.es} \affiliation{Departamento de Qu\'\i mica-F\'\i sica, Universidad del Pa\'\i s Vasco, Apdo. 644, 48080 Bilbao, Spain}
\author{G. Garc\'\i a-Calder\'on}
\altaffiliation{Permanent address: Instituto de F\'{\i}sica, Universidad Nacional Aut\'onoma de M\'exico, Apartado Postal {20 364}, 01000
M\'exico, D.F., M\'exico} \email{gaston@fisica.unam.mx} \affiliation{Departamento de Qu\'\i mica-F\'\i sica, Universidad del Pa\'\i s Vasco,
Apdo. 644, 48080 Bilbao, Spain}
\author{J. G. Muga}
\email{jg.muga@ehu.es} \affiliation{Departamento de Qu\'\i mica-F\'\i sica, Universidad del Pa\'\i s Vasco, Apdo. 644, 48080 Bilbao, Spain}
\author{M. G. Raizen}
\email{raizen@physics.utexas.edu} \affiliation{Center for Nonlinear Dynamics and Department of Physics, The University of Texas at Austin,
Austin, Texas 78712-1081, USA}

\def\la{\langle}
\def\ra{\rangle}
\def\om{\omega}
\def\Om{\Omega}
\def\vep{\varepsilon}
\def\wh{\widehat}
\def\tr{\rm{Tr}}
\def\da{\dagger}
\newcommand{\beq}{\begin{equation}}
\newcommand{\eeq}{\end{equation}}
\newcommand{\beqa}{\begin{eqnarray}}
\newcommand{\eeqa}{\end{eqnarray}}
\newcommand{\intf}{\int_{-\infty}^\infty}
\newcommand{\into}{\int_0^\infty}
\date{\today}
\begin{abstract}
We  study the tunneling dynamics of  bosonic and fermionic Tonks-Girardeau gases  from a hard wall trap, in which one of the walls is
substituted by a delta potential. Using the Fermi-Bose map, the decay of the probability to remain in the trap is studied as a function of both
the number of particles and the intensity of the end-cap delta laser. The fermionic gas is shown to be a good candidate to study deviations of
the non-exponential decay of the single-particle type, whereas for the bosonic case a novel regime of non-exponential decay appears due to the
contributions of different resonances of the trap.

\end{abstract}
%
%
\maketitle
\section{Introduction}
Decay of a metastable system via tunneling is one of the most remarkable and old effects in quantum mechanics. Since Gamow's analysis of alpha
decay, resonance theory, which applies to virtually all fields from particle to molecular physics, has been motivated by this phenomenon. Simple
treatments examine the escape or survival of single particle wave functions in one dimensional (1D) potentials. At this level much attention has
been paid to deviations from exponential decay, and Zeno or anti-Zeno effects. Also, exact results are typically available, by means of
analytical models or numerically. The more complex decay of a multiparticle unstable system is treated by more sophisticated multichannel, or
reactive-scattering approaches, sometimes with statistical approximations or, depending on the system and environment, in a phenomenological
way, and also using mean-field approximations. In ``macroscopic quantum tunneling'', a macroscopic variable, such as the phase difference of the
Cooper pair wave function across a Josephson junction
obeys a simple tunneling equation for an effective particle subjected to dissipation \cite{CL83, JJ85}. The effect of dissipation due to the
perturbing environment has thus been extensively discussed and measured. Other macroscopic quantum tunneling effect much studied in recent times
is the tunneling and decay of Bose-Einstein condensates; in particular the effect of effective atom-atom interaction and the non-linear term in
the Gross-Pitaevskii mean-field approach \cite{BECS98,BECS01,BECS05}. 

Some works go beyond the mean field theory using simplified Hamiltonians
\cite{EJWH06}, which is particularly relevant for few-body systems. 
These systems with not too many particles may still be amenable of exact treatments but will show differences from single and many-particle ones.     
The experimental study of few-body tunneling is a formidable prospect, as it requires initial preparation of a ground few-body Fock state, precise control over the tunneling time, and the ability to count single atoms with unit quantum efficiency.  Until recently, such capabilities did not exist so that any experimental tests seemed unlikely.  However recent developments now open the door for few body tunneling experiments and motivate the present work in anticipation of such results.  The starting point was the development of a novel optical box trap that confines a degenerate Bose gas, together with single-atom counting  \cite{R1,R2}.  The same box trap was used to produce number squeezing of atoms by confining a degenerate Bose gas and controlled lowering of the walls until a final value \cite{R3}.  The observed fluctuations in number were a factor of two below the Poissonian limit, but the residual noise can be accounted for by known sources of technical noise, so that these experiments are consistent with number-state production.  This simple procedure, called ``laser culling of atoms'' has recently been analyzed theoretically, and is shown to produce atomic few-body Fock states for sufficiently slow ramp time and neglecting quantum tunneling through the barrier \cite{R4}.
The latter effect can be highly suppressed by sculpting the shape of the barrier using the techniques described in \cite{R2}. The barrier width can then be reduced at a well defined time, allowing quantum tunneling to occur.  This system should therefore enable the first experimental study of few body quantum tunneling in different regimes of interaction and with controlled number.

In this paper we study the decay from a trap by tunneling trough a delta barrier of a 
few-body Tonks-Girardeau gas with
the aim of obtaining exact results of a few-body decay problem for which the experimental verification is in view \cite{R1,R2,R3}. In particular,
we find few-body deviations from the exponential decay law. For one particle, deviations from exponential decay have been long predicted at both short and long times.
Short time deviations were observed experimentally with ultracold atoms \cite{W97}, whereas long time deviations have been
observed very recently for the first time in dissolved organic materials \cite{R06}. A suitable system to test deviations from exponential decay at the
few-body level is the bosonic Tonks-Girardeau (BTG) gas \cite{Girardeau60}. Such a gas actually mimics the fermionic behavior to minimize
the strongly repulsive interaction. At low densities, the TG regime can be reached under a strong enough radial confinement \cite{Olshanii98} such that the
transverse degrees of freedom are reduced to zero-point oscillations, resulting a 1D effective system. Indeed, experimental observations have
already been reported \cite{exp}.
%

As another striking example of the Fermi-Bose duality \cite{CS99}, the fermionic Tonks-Girardeau (FTG) gas has been described as a 1D spin-aligned
fermionic gas in the limit of highly attractive interactions mediated by a 3D $p$-wave Feshbach resonance \cite{ftg}. The FTG gas is in a sense
the opposite of the BTG gas and exhibits ``bosonization'' because of the strong attraction.

In this work we study the tunneling dynamics of both bosonic and fermionic TG gases, initially confined in a hard wall potential. For the bosonic
case, the Fermi-Bose (FB) map \cite{Girardeau60,YG05,CS99,GW00b} relates the  wavefunction of $N$ strongly interacting bosons,
$\psi_{B}$, to the one of an ideal Fermi gas with all spins frozen in the same direction, $\widetilde{\psi}_{F}$, where the tilde indicates the
``dual'', or auxiliary system. The Fermi wavefunction, is built as a Slater determinant,
\beq \widetilde{\psi}_{F}(x_{1},\dots,x_{N}) =\frac{1}{\sqrt{N!}}det_{n,k=1}^{N}\phi_{n}(x_{k}), \eeq
where $\phi_{n}(x_{k})$ are the eigenstates of the trap. To account for the proper quantum statistics, the ``antisymmetric unit function"
\beq \mathcal{A}=\prod_{1\leq j<k\leq N}sgn(x_{k}-x_{j}) \eeq
is introduced in such a way that
\beq \psi_{B}(x_{1},\dots,x_{N})= \mathcal{A}(x_{1},\dots,x_{N})\widetilde{\psi}_{F}(x_{1},\dots,x_{N}). \eeq
In so doing, $\psi_{B}(x_{1},\dots,x_{N})$ becomes totally symmetric under permutation of particles. Moreover, since  $\mathcal{A}$ is
involutive and does not include time explicitly, for any unitary evolution it can be proved that the time-dependent density profile can be
calculated as \cite{GW00b}
\begin{eqnarray}
\rho_{B}(x,t)&= &N\!\!\int\vert\psi_{B}(x,x_{2},\dots,x_{N};t)\vert^{2}dx_{2} \cdots dx_{N} \nonumber \\ [.4cm]
&=&\sum_{n=1}^{N}\vert\phi_{n}(x,t)\vert^{2}. \label{11}
\end{eqnarray}
In a similar way, one can deal with the Fermionic TG (FTG) gas \cite{ftg}, but using now the generalized FB mapping in the opposite direction.
The wavefunction of the FTG gas can be written in terms of the Hartree product describing the dual (auxiliary) system, which is now
the ideal Bose gas,
\beq \widetilde{\psi}_{B}(x_{1},\cdots,x_{N}) =\prod_{i=1}^{N}\phi_{0}(x_{i}). \eeq
The proper symmetrization is carried out also through the antisymmetric unit function,
\beq \psi_{F}(x_{1},\cdots,x_{N})=\mathcal{A}(x_{1},\dots,x_{N}) \widetilde{\psi}_{B}(x_{1},\dots,x_{N}), \eeq
in such a way that the density profile can be written as
\begin{equation}
\rho_{F}(x,t)=N\vert\phi_{0}(x,t)\vert^{2}. \label{12}
\end{equation}

The trap we will consider is of hard-wall kind after its experimental realization in its all-optical \cite{MSHCR05} or  microelectronic chip
version \cite{HHHR01}. These traps have recently generated much theoretical work in the field of ultracold atoms in low dimensions
\cite{Gaudin71,Cazalilla02, BGOL05,DM05b}.

\section{Single eigenmode dynamics}
First of all we shall study the time evolution of the $n$-th eigenstate of a hard wall trap. As it is well-known they have the general form
\begin{equation}
\phi_{n}(x,t=0)=\sqrt{\frac{2}{L}}\sin \left(\frac{n\pi x}{L}\right )\chi_{[0,L]}(x), \label{1a}
\end{equation}
with $n\in\mathbb{N}$ and $\chi_{[0,L]}(x)$ the characteristic function in $[0,L]$.

At time equal zero the right-wall is substituted by a delta potential, $V(x)=\eta\, '\delta(x-L)$,  which is to represent a far-detuned laser from
the atomic resonance, so that atomic excitation becomes negligible. This model has been recently considered to study non-exponential decay at
both short and long times at the single-particle level in \cite{MDDG05}. The time evolution of a given initial state may be written in terms of
the retarded Green's function $g(x,x';t)$ as
\begin{equation}
\phi_n(x,t) = \int_0^L g(x,x';t)\phi_n(x',0) dx'. \label{1}
\end{equation}
Equation (\ref{1}) may be calculated using an expansion in the eigenfunctions of the Hamiltonian $\{\vert k^{+}\ra\vert k\in\mathbb{R}^{+}\}$,
\textit{i.e.}, the so called \textit{physical} wave solutions,
\beqa 
\phi_{n}(x,t)&=&\int_{0}^{\infty}dk\la x\vert k^{+}\ra \la k^{+}\vert\phi_{n}\ra e^{-i\hbar k^{2}t/2m},
\\
%
%
\la x\vert k^{+}\ra &=& \sqrt{\frac{2}{\pi}} \left \{
\begin{array}{cc}
\sin(kx)/J_+(k),\;\; x\leq L \\[.5cm]
(i/2)\left[e^{-ikx}- S(k) e^{ikx}\right] & ,\;\; x \geq L,
\end{array}
\right. \label{2}
\end{eqnarray}
where the S-matrix $S(k)=J_{-}(k)/J_+(k)$, with $ J_-(k)=J_+^*(k)$,  and the Jost function $J_+(k)$ reads
\begin{equation}
J_+(k)=1+ \frac{\eta}{2ik}( e^{2ikL}-1). \label{3}
\end{equation}
Above and in the rest of this work we take $k=[2mE/\hbar^2]^{1/2}$ and $\eta=[2m/\hbar^2]\eta\, '$, $E$ being the energy of the decaying particle
and $m$ its mass.

The position of the pole $k_j$ of the $S$-matrix in the $k$-plane gives us information about the lifetime $\hbar/\Gamma_{j}$ of the resonance
and its real energy $\mathcal{E}_{j}$ in the leaking trap (see Fig. 1): $E_j=k_{j}^2\hbar^2/(2m)=\mathcal{E}_{j}-i\Gamma_{j}/2$. We notice that
for a weaker laser (smaller $\eta$), the resonance decays faster. Besides, $j$ is chosen so that the width of the resonance increases
monotonically with it, $j=1$ corresponding to the longest lived resonance.

The limit $\eta\rightarrow0$ is an important reference corresponding to free evolution in the presence of a wall at the origin, once the endcap
laser at $x=L$ has been turned off. In such case, a fully analytical solution is available invoking the method of images \cite{Kleber94}, see a detailed study of the BTG expansion dynamics in the
absence of the wall in \cite{DM05b}.
%
%
%
%
%
%
%
%
%

\begin{figure}[!tbp]
\includegraphics[width=2.6in, angle=-90]{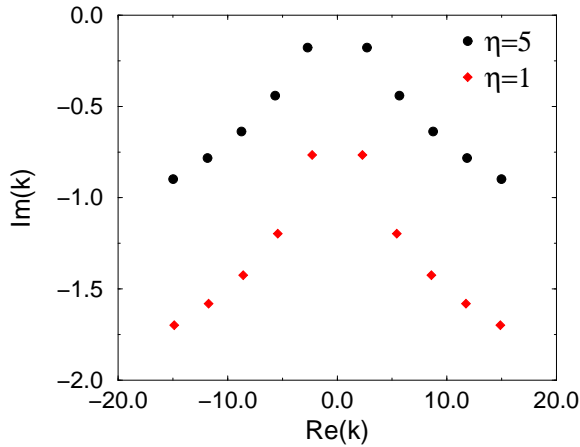}
\caption{\label{poles} Diagram on the complex $k$-plane of the first five resonance and anti-resonance poles of the hard-wall leaking trap at
two different strengths of the end-cap laser.
Here and in all figures we use dimensionless units with 
$L=1$ and $2m=\hbar=1$.} 
\end{figure}
%
%

%
\begin{figure}
\includegraphics[width=3.3in]{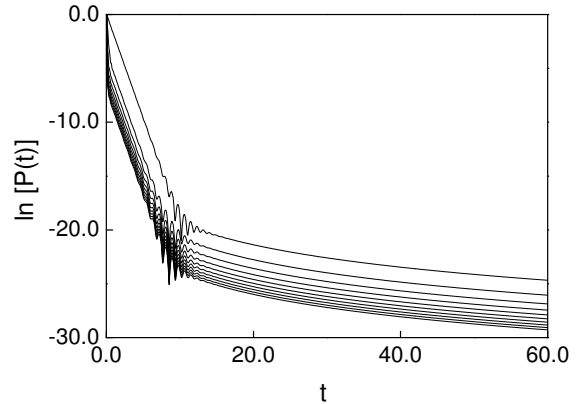}
\caption{\label{ndeltalong} Single particle nonescape probability exhibiting the long time nonexponential behavior for the first ten
eigenstates of the hard wall trap, tunneling through a delta potential with $\eta=5$. From top to bottom: $n=1,\dots,10$.}
\end{figure}
%

\begin{figure}
\includegraphics[width=0.778\linewidth, angle=-90]{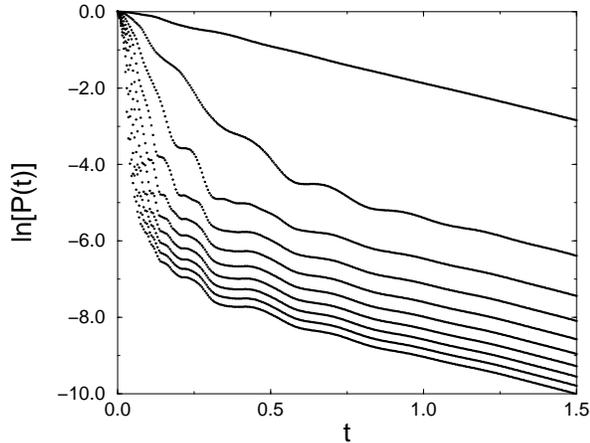}
\caption{\label{ndelta} Single particle nonescape probability for the first ten eigenstates of the hard wall trap, tunneling through a delta
potential with $\eta=5$ at short times. From top to bottom: $n=1,\dots,10$.}
\end{figure}

The decay of the different eigenstates of the hard wall trap is shown in Fig. \ref{ndeltalong}. It  exhibits the characteristic transition to
long times as an inverse power of time. The scale of the figure conceals an interesting behavior at short times, which is exhibited in Fig. \
\ref{ndelta}. We observe, with exception of the first decaying state, that each state presents some  characteristic oscillations at short times
and then tends to decay with the same slope as the first state. Physical insight regarding this behavior may be obtained making use of the
formalism of resonant states. In this representation one Laplace transforms $g(x,x';t)$ in Eq.\ (\ref{1}) into the complex \textit{k}-plane to
exploit the analytical properties of the corresponding outgoing Green's function $G^+(x,x';k)$ \cite{GMM95}. This leads to an alternative
expression to Eq.\ (\ref{2}), namely,
\begin{eqnarray}
&&\phi_n(x,t)= \sum_{j=1}^{\infty} c_j(n) u_j(x)e^{-i\hbar k_j^2t/2m} + \nonumber \\ [.4cm] &&\frac{i}{\pi}\int_0^L \phi_n(x',0)dx'  \int_{C_L}
G^+(x,x';k)e^{-i\hbar k^2 t/2m} k dk,\nonumber \\
 \label{4}
\end{eqnarray}
where the sum runs over the so called proper complex poles \textit{i.e.}, those on the fourth quadrant of the complex $k$-plane with ${\rm Re}(k_j)>|{\rm Im} (k_j)|$;
the $u_j$'s obey outgoing boundary conditions at $x=L$ and satisfy the Schr\"odinger equation with complex eigenvalues $E_j=\mathcal{E}_j-i\Gamma_j/2$;  the
coefficients $c_j(n)$ give the overlap of the initial state $\phi_n$ with the resonant states $u_j$ of the problem, namely,
$c_j(n)=\int_0^L\phi_n(x,0)u_j(x)dx$;  the integral term  involving $G^+(x,x';k)$ stands for the non-exponential contribution, which in general
may be neglected except at ultrashort or very long times. The path  $C_L$ of the integral is chosen, without loss of generality, as a straight
line 45 degrees off the real \textit{k}-axis along the complex $k$-plane passing through the origin $k=0$ \cite{GMM95}.

%
%

The nonescape probability is defined as
\begin{equation}
P_n(t)=\int_0^L  |\phi_n(x,t)|^2 dx,
\label{5}
\end{equation}
(to be distinguished from the ``survival probability'' 
$|\la \phi_n(t)|\phi_n(0)\ra|^2$, which is harder to measure).
Using the exponential contribution to Eq.\ (\ref{4}) into Eq.\ (\ref{5}) yields the nonescape probability for each decaying eigenstate as
\begin{equation}
P_n^{\,exp}(t)= \sum_{s,j}^Nc_j(n)c_s^*(n) I_{js}e^{-i(\mathcal{E}_j-\mathcal{E}_s)t/\hbar} e^{-(\Gamma_j+\Gamma_s)t/2\hbar},
\nonumber\\
\label{nonscp}
\end{equation}
where $I_{js}=\int_0^Lu_j(x)u_s^*(x)dx$. The exponential term in Eq.\ (\ref{nonscp}), with $N=n$, reproduces very well the regime depicted in
Fig. \ref{ndelta}. It reflects a transient regime where each eigenstate $\phi_n$ makes eventually a transition into the longest lived eigenstate
($j=1$). As shown in Fig. \ref{ndeltalong}, after the exponential regime, it follows the long time $t^{-3}$ inverse power law governed by the
integral contribution to Eq.\ (\ref{4}). 
Standard asymptotic analysis gives the result 
%
%
%
%
%
\beq 
P_{n}^{\,long}(t) = \left( \frac{2m}{\hbar}\right )^3 \frac{L^{3}}{12\pi(1+\eta L)^{4}}\frac{\mathcal{C}^{2}(n)}{t^{3}},  \label{longt}
\eeq
where $\mathcal{C}(n)=\int_{0}^{L}\phi_{n}(x,0)xdx=L\sqrt{2L}(-1)^{n}/(n\pi)$. Clearly the transients in the transition from exponential to
nonexponential behavior observed in Fig. \ref{ndeltalong} originate from the interference between the exponential and long time expressions of
$\phi_n(x,t)$.
The transition may be displaced to earlier times and made more easily observable 
by decreasing the ratio $R=\mathcal{E}_1/\Gamma_1$ of the 
longest lived resonance \cite{Jittoh,GCV}.

%
%
%

{\section{Tonks-Girardeau gas}}
We next generalize the notion of nonescape probability to a few particle system, associating it with the average number of particles within
the trap at a given time,
\beq
N_T(t)=\int_{0}^{L}dx\rho(x,t),
\eeq
where $\rho(x,t)$ is the density normalized to the total number of particles $N$. The nonescape probability per particle thus becomes
$P(t)=N_T/N$. From this definition it is clear that for the FTG gas, $P(t)$ is identical to the single particle nonescape probability
associated with the ground state of the trap. However, this means that the total signal is enhanced by a
factor corresponding to the number of particles, a key advantage for the experimental study of deviations from the exponential decay law. In
other words, $\ln N_T$ for an $N$-particle FTG gas in its ground state, is obtained by shifting upwards the curve for $n=1$ in Figs.
\ref{ndeltalong} or \ref{ndelta} by $\ln N$.

\begin{figure}
\includegraphics[height=5cm]{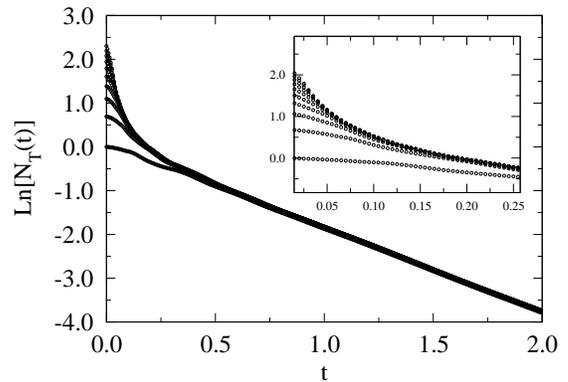}
\caption{\label{mode} Logarithm of the average number of particles within the trap for a BTG gas as a function of time, for different values of
the total number of particles. From top to bottom, $N=1,\dots,10$ with $\eta=5$. The curve for the FTG gas of $N$-particles is obtained by
shifting up the $N=1$ BTG gas
curve (single-particle case) by $\ln N$.
}
\end{figure}

More remarkable yet, the BTG gas exhibits novel few-body decay features. In particular, a new regime of non-exponential decay arises from the
sum of contributions of the single particle exponentials. Figure \ref{mode} shows that, for the BTG gas, the higher the number of particles the
faster is the tunneling rate and, as a result of the different energy contributions, the short time dependence is affected.
%
%

The effect of the end-cap laser intensity can be simulated by varying $\eta$. In so doing,  we have learnt that the resonance poles shift
towards the real axis in the $k$-plane for increasing $\eta$, see Fig. \ref{poles}. For vanishing $\eta$, even for the single particle case, and
therefore the FTG gas, there is no reason to expect exponential decay since the potential is not a trap any more and there are no resonances.

%
\begin{figure}
\includegraphics[width=2.8in, angle=0]{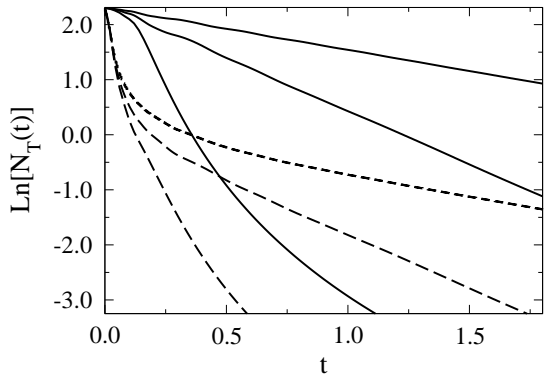}
\caption{\label{eta} Logarithm of the average number of particles in the trap for a BTG gas (dashed lines) and FTG gas (continuous lines) of 10
particles for different intensities of the end-cap laser (for both cases, from top to bottom $\eta=10,5,0$).}
\end{figure}
%
This behavior is shown in Fig. \ref{eta}. For a given $\eta$, a clear difference between BTG and FTG gases is the decay rate, which is larger
for the former at short times, a direct consequence of the different ground state energy for both systems, namely,
$$
E_{BTG}=\hbar^{2}\pi^{2}N(N+1)(2N+1)/(12mL^2)
$$
and
$$
E_{FTG}=\hbar^{2}\pi^{2}N/(2mL^2).
$$
However, the high resonance contributions eventually
diminish, and the rates become finally equal being governed by the longest lived resonance, deep into the purely (single) exponential
regime. Notice that at short times, the decay is dominated by high energy components and it is essentially independent of the strength of the
delta potential. Moreover, at short times the non-exponential deviation is concave-up for the BTG system and concave-down for the fermionic
case.

Regarding the long time behavior, it follows from Eq.\ (\ref{longt}) that $N_T(t)=NP_0(t)$ for the FTG gas exhibits the $1/t^{3}$ dependence.
This is also the case for BTG gas where \cite{explain}
\begin{equation}
N_T(t)=\sum_{n=1}^{N}P_{n}(t)\propto\sum_{n=1}^{N}\frac{\mathcal{C}^{2}(n)}{t^{3}}. \label{13}
\end{equation}
%
This result guarantees that the generalization for our few-body systems of the dwell time (the average time spent by a particle in a spatial
region) is possible, being a meaningful finite quantity \cite{MDS95}.\vspace*{.5cm}
%
%

%
In conclusion, we have studied exactly two related few-body tunneling problems, namely, that of  bosonic and fermionic TG gases. The FTG gas
has been pointed out as a good system to observe long time deviations from exponential decay because of the strength of the signal, proportional
to the number of particles. The recent first measurement of long time deviations \cite{R06} is based on the important effect of the environment
in organic molecules in solution, so that the deviation for ``pure'', isolated  systems 
remains to be observed \cite{GCV}. For the bosonic case a new
deviation of the exponential decay appears, which can be understood as a sum of $N$ single-particle contributions.

It is still an experimental challenge to get to the strong TG limit in a flat box, the main limiting factor being the confinement in transverse
directions. Assuming, according to Reichel and Thywissen \cite{RT}, a ``maximal practical value'' of transversal frequency of 1 MHz, $N=10$, a
box of $10 \mu$m, and the constants for rubidium 87 (scattering lenght $a\approx 5$ nm), the ratio $\alpha$ of interaction energy to the
potential energy is $\alpha\approx 440$, whereas the ratio of chemical potential to the kinetic energy is $\gamma\approx 88$, see \cite{RT}. For rubidium 85 at a 
Feshbach resonance the scattering length, $\alpha$ and $\gamma$  may increase by a factor of $100$ \cite{85}, making optical traps a viable alternative to magnetic confinement. These values point
out at a TG gas regime not too far away from current capabilities, but the actual implementation may still be difficult also because of the need
for accurate single atom detection. 

We may in any case expect that reaching a strict
TG regime is not absolutely essential to find interesting few-body decay effects.   
The upshot of our analysis is that for the few-body ground state of the box,
the BTG gas exhibits maximal deviations from the exponential law at short times, whereas they are completely absent in the FTG gas or ideal Bose gas. 
The behaviour between the ideal Bose gas and BTG gas could be smoothly extrapolated by increasing the effective interactions \cite{chinos}. 
Moreover, the Bose-Fermi duality \cite{CS99} opens up the possibility of observing experimentally such effects in a wider class of systems.\vspace*{.6cm} 
\begin{acknowledgments}

This work has been supported by Ministerio de Educaci\'on y Ciencia (BFM2003-01003), and UPV-EHU (00039.310-15968/2004). A. C. acknowledges
financial support by the Basque Government (BFI04.479) and G. G-C.
the financial support of El Ministerio de Educaci\'on y Ciencia (SAB2004-0010).
M. G. R. acknowledges support from NSF, the R. A. Welch
Foundation, and the S. W. Richardson Foundation and the US Office of Naval
Research, Quantum Optics Initiative, Grant N0014-04-1-0336.\\
\end{acknowledgments}
%


\begin{thebibliography}{10}
\bibitem{explain} The time scale for which the asymtotic form in Eq. \ref{longt} occurs is given by its intersection with the exponential term associated with the first resonance.
\end{thebibliography}
\end{document}